\documentstyle[preprint,epsfig,aps]{revtex}
\input epsf.tex
\def\DESepsf(#1 width #2){\epsfxsize=#2 \epsfbox{#1}}
\begin{document}
\preprint{\vbox{\hbox{KAIST-TH 2001/08}}}
\draft
\title{
Muon Anomalous $g -2$ and Gauged $L_\mu - L_\tau$ Models
}
\author{S. Baek$^1$, N.G. Deshpande$^2$, X.-G. He$^1$ and P. Ko$^3$}
\address{
$^1$Department of Physics, National Taiwan University, 
Taipei, Taiwan\\
$^2$ Institute of Theoretical Sciences, University of Oregon, Eugene,
OR 97402\\
$^3$   Department of Physics, KAIST, Taejon 305-701, KOREA          
}

\tighten

\date{ April, 2001}
\maketitle
\begin{abstract}
In this paper we study $Z'$ contribution to  $g -2$ of the muon
anomalous  magnetic dipole moment in gauged  $U(1)_{L_\mu - L_\tau}$
models. Here $L_i$ are the lepton numbers.  We find that there are
three classes of models which can produce a large value of $g-2$ to
account for possible discrepancy between the experimental data  and
the Standard Model prediction. The three classes are:  a) Models with
an exact $U(1)_{L_\mu - L_\tau}$.  In these models, $Z'$ is
massless. The new gauge interaction coupling $e a/\cos\theta_W$ 
is constrained to
be  $ 0.8\times 10^{-3} < |a| < 2.24\times 10^{-3}$.  b) Models with
broken $U(1)_{L_\mu - L_\tau}$ and the breaking scale is not related
to electroweak symmetry breaking scale. The $Z'$ gauge boson is
massive.  The allowed range of the coupling and the $Z'$ mass are
constrained, but  $Z'$ mass can be large; And c)  The
$U(1)_{L_\mu-L_\tau}$ is broken and the breaking scale is related to
the electroweak scale. In this case the $Z'$ mass is constrained to
be $\sim 1.2$ GeV. We find that there are interesting experimental
signatures in  $\mu^+\mu^-\to \mu^+\mu^-, \tau^+\tau^-$ in these
models.
\end{abstract}


\preprint{\vbox{\hbox{}}}

\section{Introduction}

Recently experiment from BNL\cite{bnl} has measured 
the muon anomalous magnetic 
dipole moment with $a_\mu^{exp} = 
(g -2)/2=(11 659 202 \pm 14\pm 6) \times 10^{-10}$.
This value differs the Standard Model (SM) prediction 
in Ref.\cite{dh,marciano}  
by 2.6$\sigma$,

\begin{eqnarray}
\Delta a_\mu = a^{exp}_\mu - a^{SM}_\mu = (42.6\pm 16.5) \times 10^{-10}.
\label{eq:BNL}
\end{eqnarray}

At present the experimental errors are still too large 
to claim a real deviation. There are also uncertainties from 
theoretical calculations, in particular contributions from hadrons at loop
levels are not well determined~\cite{yndurain}. 
Improvements from both experimental measurements
and theoretical calculations are needed.
If this difference is true, it is an indication of new physics beyond the
SM. Many authors have discussed possible implications for new physics
beyond the SM~\cite{bnlanal}. 
Some interesting constraints have been obtained.
In this paper we study the implications of a large $\Delta a_\mu$
on models with gauged $L_\mu - L_\tau$. Here
$L_i$ is the $i$ lepton number.

$L_\mu - L_\tau$ gauge models are some of the simplest models beyond the
SM which contain an additional $Z'$ boson. 
Without enlarging the fermion contents in the SM, there are only 
three types of U(1) symmetries which can be gauged from anomaly cancellation 
requirement. These symmetries are

\begin{eqnarray}
i)\;\; U(1)_{L_e -L_\mu};\;\;\;\;ii)\;\; U(1)_{ L_e -L_\tau};
\;\;\;\;iii)\;\; U(1)_{ L_\mu - L_\tau}.
\end{eqnarray} 

Some experimental consequences of these 
models have been studied in Refs.\cite{he,he1}. 
There are stringent constraints on the parameters of models based
on i) and ii) because the $Z'$ couple to electrons. 
It is difficult to generate a large enough value
for $\Delta a_\mu$ in eq. (1). On the other hand, for models based on iii)
there are limited data available to constrain relevant parameters. 
It is possible
to have a large $\Delta a_\mu$. 

In $U(1)_{L_\mu-L_\tau}$ models, only the second and third generations of
leptons are affected, whereas all other
SM particles are not. 
The transformation properties of leptons under the $SU(3)_C\times SU(2)_L
\times U(1)_Y$ SM gauge group 
and the $U(1)_{L_\mu - L_\tau}$ gauge group are

\begin{eqnarray}
\begin{array}[1]{clcl}
L^e_L: & (1,2,-1)(0) & e_R: & (1,1,-2)(0) \\
L^\mu_L: & (1,2,-1)(2a) & \mu_R: &(1,1,-2)(2a) \\
L^\tau_L: & (1,2,-1)(-2a) \quad & \mu_R: & (1,1,-2)(-2a).
\end{array}
\end{eqnarray}
where the numbers in the first and the second brackets indicate 
the transformation
properties under the SM gauge group and the $U(1)_{L_\mu - L_\tau}$
group, respectively. The numbers in the second bracket will be indicated
as $Y'$. 
The covariant derivative in terms of the photon
field $A_\mu$, the $Z_\mu$ field, and the $Z'_\mu$ field is given as

\begin{eqnarray}
D_\mu = \partial_\mu + ie QA_\mu + i {e\over s_W c_W} (I_3-s^2_W Q) Z_\mu
+ i {e\over c_W} {Y'\over 2} Z'_\mu,
\end{eqnarray}
where $s_W = \sin \theta_W$, $c_W = \cos\theta_W$. We have normalized
the $Z'$ coupling to the $U(1)_Y$ charge coupling $e/c_W$.

The $U(1)_{L_\mu - L\tau}$ may be an exact symmetry
or broken at some scale which may or may not be related to the electroweak
breaking scale. One can classify three types of models based on
$U(1)_{L_\mu - L_\tau}$ as:
a) $U(1)_{L_\mu - L_\tau}$ is an exact symmetry; 
b) $U(1)_{L_\mu - L_\tau}$ is broken and the breaking 
scale is not related to the
electroweak scale; And
c) $U(1)_{L_\mu - L_\tau}$ is broken and the breaking scale is related to
the electroweak scale. In all these cases $\Delta a_\mu$ receives contribution 
from $Z'$ exchange at one loop level. In case c), there is in general $Z-Z'$
mixing. Electroweak precision tests from various experiments constrain the 
mixing severely. We will concentrate on models where the $Z-Z'$ mixing is
naturally zero at tree level, such that the related constraints are 
automatically satisfied.

\section{The muon magnetic dipole moment in $U(1)_{L_\mu - L_\tau}$ Models}

\noindent
{\bf $\Delta a_\mu$ in case a)}

In this case, there is no need to introduce other new particles. It is the
simplest $U(1)_{L_\mu - L_\tau}$ model.
$Z'$ couplings to $\mu$ and $\tau$ are given by

\begin{eqnarray}
L = {ea\over c_W} (\bar \mu \gamma^\mu \mu - \bar \tau \gamma^\mu \tau) Z'_\mu.
\end{eqnarray}
The Feynman diagram which generates a non-zero $\Delta a_\mu$ is shown
in Fig. 1.
The contribution to $\Delta a_\mu$ is given by

\begin{eqnarray}
\Delta a_\mu = a_\mu^{SM+new} - a_\mu^{SM} = 
{\alpha_{em}\over 2\pi}
{a^2\over c^2_W}.
\end{eqnarray}
The 2$\sigma$ range consistent with eq.~(\ref{eq:BNL}) is determined to
be
\begin{eqnarray}
0.8\times 10^{-3} < |a| < 2.24 \times 10^{-3}.
\label{eq:casea}
\end{eqnarray}
The $\tau$ also receives the same amount of correction to the anomalous 
magnetic dipole moment.

With a non-zero value for $a$, all processes involving $\mu$ and $\tau$ 
will be affected. 
However, because the coupling parameter 
$a$ is small, the effects are all small.

\noindent
{\bf $\Delta a_\mu$ in case b)}

The simplest model for case b) can be realized by just introducing a SM 
singlet scalar $S$ but transforms non-trivially under $U(1)_{L_\mu - L_\tau}$, 
$S:\;\;(1,1,0)(a)$. 
In this case when $S$ develops a non-zero vacuum expectation value (VEV) 
$v_S$, the $Z'$ boson becomes massive with 
$m^2_{Z'} = e^2 a^2 v_S^2/2 c^2_W$. In this model, there is no $Z$ 
and $Z'$ mixing at tree level. The $Z'$ couplings to
$\mu$ and $\tau$ are the same as that in eq. (5). One obtains a non-zero 
$\Delta a_\mu$ through the same diagram in Fig. 1 for case a), but with 
a non-zero $Z'$ mass. We have
\begin{eqnarray}
\Delta a_\mu = {\alpha_{em} \over 2\pi}{a^2\over c^2_W}
  \int_0^1 dx \frac{2 m^2_\mu x^2 (1-x)}{x^2 m^2_\mu +(1-x) m_{Z'}^2}.
\label{eq:g2_b}
\end{eqnarray}

Requiring the new contribution to produce the value in eq.~(\ref{eq:BNL}), the 
allowed values of $a$ and $m^2_{Z'}$ are constrained. The 
results are shown in Fig.~2. We see that there is a large allowed region where
a large value of $\Delta a_\mu$ can be produced. 

In the limit $m^2_{Z'} >> m^2_\mu$,
\begin{eqnarray}
\Delta a_\mu = {\alpha_{em} \over 2 \pi} {a^2\over c^2_W} 
{2\over 3}{m^2_\mu \over m^2_{Z'}} .
\end{eqnarray}
To produce the value in eq. (1), one obtains 
$9.2\times 10^{-3} <a/m_{Z'} ({\rm GeV}) < 25.8 \times 10^{-3}$. 
The breaking scale $v_S$ of the
$U(1)_{L_\mu - L_\tau}$ is of order $\sim 200$ (GeV). 
Changing $m_\mu$ to $m_\tau$ in eq. (\ref{eq:g2_b}), one obtains the tauon
$g-2$.
We note that for large enough $m_{Z'}$ only the parameter $a/m_{Z'}$
is constrained from $\Delta a_\mu$. Of course one should not let $a$ to
be arbitrarily large, because a large $a$ will invalidate perturbation
calculations carried out here. We will limit $a$ such that
$\alpha_{em} a^2 \lesssim 1$.
The effects of $Z'$ on
$\mu^+\mu^-\to \mu^+\mu^-, \tau^+ \tau^-$ turn out to be quite dramatic
in this case and will be discussed in the next section.

\noindent
{\bf $\Delta a_\mu$ in case c)}

There are many ways to realize case c). Here we study the effect on 
$\Delta a_\mu$ in the model discussed in Ref.\cite{he1}. 
In this model there are 
two more SM Higgs doublets 
$\phi_{2,3}$ in addition to the 
usual SM doublet $\phi_1$.
The Higgs doublets 
SM gauge group and the $U(1)_{L_\mu -L_\tau}$ quantum numbers are

\begin{eqnarray}
\phi_1 : (1,2,1)(0);\;\;\phi_2 : (1,2,1)(4a);\;\;\phi_3: (1,2,1)(-4a).
\end{eqnarray}
Because $\phi_{2,3}$ transform non-trivially under the SM and 
$U(1)_{L_\mu - L_\tau}$, in general after symmetry breaking
there are $Z$ and $Z'$ mixing. 
This mixing can be eliminated by applying a unbroken discrete symmetry,

\begin{eqnarray}
&&A_\mu \to A_\mu;\;\;
 Z_\mu \to Z_\mu;\;\; 
 Z'_\mu \to -Z'_\mu; \nonumber\\
&& L^\mu_L (\mu_R) \leftrightarrow L^\tau_L (\tau_R);\;\;
\phi_2 \leftrightarrow \phi_3; \nonumber\\
&&<\phi_2>=v_2=<\phi_3> = v_3.
\end{eqnarray}
The $Z'$ mass in this model is given by

\begin{eqnarray}
m^2_{Z'} = 16{e^2\over c^2_W} a^2 v^2_2.
\label{eq:zprime_b}
\end{eqnarray}

The Yukawa couplings of the $\phi_{2,3}$, consistent with the discrete 
symmetry, are given by

\begin{eqnarray}
L_{Yuk} = \lambda (\bar L^\mu_L \mu_R \phi_1 + \bar L^\tau_L \tau_R\phi_1)
+ \lambda' (\bar L^\mu_L \tau_R \phi_2 + \bar L^\tau_L \mu_R\phi_3).
\end{eqnarray}
The above Yukawa coupling produces 
a non-diagonal mass matrix for $\mu$ and $\tau$. 
In the 
mass eigenstate bases, $Z'$ couplings to $\mu$, $\tau$, and their 
associated 
neutrinos are given by

\begin{eqnarray}
L = {ea\over \cos\theta_W} [\bar \mu \gamma^\mu \tau 
+ \bar \tau \gamma^\mu \mu] Z'_\mu
+ {ea\over 2\cos\theta_W}[\bar \nu_\mu \gamma^\mu (1-\gamma_5) \nu_\tau +
\bar \nu_\tau \gamma^\mu (1-\gamma_5) \nu_\mu]Z'_\mu.
\end{eqnarray}
 
There are very stringent constraints on this model. Firstly, $a/m_{Z'}$ is 
restricted from the expression of the $Z'$ mass formula in 
eq. (\ref{eq:zprime_b}) since
$v_2$ have to be less than $\sqrt{v^2_1+v_2^2+v_3^2} = 246$ GeV which 
determines the $W$ boson mass. We have

\begin{eqnarray} 
{a^2\over m^2_{Z'}} >{1\over 16 \tan^2\theta_W m_W}.
\label{mass_c}
\end{eqnarray}

Secondly, there is new 
contribution to
$\tau^- \to \mu^- \bar \nu_\mu \nu_\tau$ by exchanging $Z'$ with
\begin{eqnarray}
R &\equiv& {\Gamma(\tau^-\to \mu^- \bar \nu_\mu \nu_\tau 
(\bar \nu_\tau \nu_\mu))\over
\Gamma(\tau^-\to \mu^-\bar \nu_\mu \nu_\tau)_{SM}}
= 1+\xi g_1(z) + 2 \xi^2 g_2(z),\nonumber\\
g_1(z) &=& -{1 \over 3z^4} \bigg[
   z(12-12z-5z^2)+6(2-3z+z^3)\log|1-z| \bigg], \nonumber\\
g_2(z) &=& {1 \over z^4} \bigg[
   z(6-3z+z^2)+6(1-z)\log|1-z| \bigg],
\end{eqnarray}
where $z = m^2_\tau/m^2_{Z'}$  and $\xi = 2\sqrt{2}\pi 
\alpha_{em}/(G_F m^2_{Z'})(a^2/ c^2_W)$. 
The factor 2 in front of $g_2(z)$ comes from the fact that
$\tau^-\to \mu^- \bar \nu_\tau \nu_\mu$ is not distinguished from
$\tau^-\to \mu^- \bar \nu_\mu \nu_\tau$ in experiments, and
we need to sum over these two modes.

Experimentally 
the SM prediction is very close to the observation for
$\tau^-\to \mu^- \bar \nu_\mu \nu_\tau$. The new contribution must 
be smaller than the experimental error\cite{pdg} 
on $R$, $\Delta R = 0.004 (1\sigma)$.
This provides a very tight constraint on the allowed parameters.

Finally there is a constraint from $\Delta a_\mu$. The Feynman diagram 
generating a non-zero $\Delta a_\mu$ is similar to Fig. 1 with the replacement
of $\mu$ by $\tau$ for the fermion in the loop. We have

\begin{eqnarray}
\Delta a_\mu &=& {\alpha_{em} \over 2\pi}{a^2\over c^2_W} 2 m_\mu\nonumber\\
 & \times&
  \int_0^1 dx \frac{x (1-x) 
 \left\{2 (m_\tau-m_\mu)+m_\mu x\right\}
 -{1 \over 2} \left(m_\tau -m_\mu \over m_{Z'}\right)^2 x^2
  (m_\tau -m_\mu +x m_\mu )}
{x^2 m^2_\mu +(1-x) m_{Z'}^2+x(m_\tau^2-m_\mu^2)}.
\end{eqnarray}

In the limit $m_{Z'} >> m_\tau$, 

\begin{eqnarray}
\Delta a_\mu = {\alpha_{em}\over 2 \pi} {a^2\over c^2_W} 
{2m_\mu m_\tau\over m^2_{Z'}}.
\end{eqnarray}

The above constraints are so restrictive that within the experimentally allowed 
value for $R-1$ and the constraint of 
eq.~(\ref{mass_c}), it is not possible to produce 
$\Delta a_\mu$ given in eq. (1). This simple model is ruled out.
 
The above problem, however, can be easily overcome by lifting the constraint 
from eq.~(\ref{mass_c}). 
This can be achieved by introducing a SM singlet $S$ for 
case b). The introduction of this singlet scalar does not cause 
$Z-Z'$ mixing and does not change the $Z'$ couplings to $\mu, \tau$ and 
their associated neutrinos, but will introduce a new contribution to 
the $Z'$ mass. The new $Z'$ mass is given by

\begin{eqnarray}
m^2_{Z'} =16{e^2a^2\over c^2_W} v^2_2 + {1\over 2}{e^2a^2\over c^2_W} v_S^2.
\end{eqnarray}
Because $v_S$ is not fixed, the constraint on $a/m_{Z'}$ from eq.~(\ref{mass_c})
is no longer applicable.

In this modified model, it is possible to obtain a large enough value for 
$\Delta a_\mu$ in eq. (1). However, the allowed parameter space is still 
very restrictive. The results are shown in Fig. 3 and 4. In Fig.3 we show the
allowed region of $a$ and $m_{Z'}$ and in Fig.4 we show the allowed 
$\Delta a_\mu$ as a function of $m_{Z'}$. To produce a large enough 
$\Delta a_\mu$ to account for the value in eq. (1), the $Z'$ mass is 
forced to be around 1.2 GeV. Note that the region $m_{Z'} < 0.5$ GeV
is ruled out by the non-observation of two body decay mode
$\tau \to \mu Z'$~\cite{he1}.

$\tau$ anomalous magnetic dipole moment also receives a similar correction. 
In the heavy $Z'$ limit, $\Delta a_{\tau} = \Delta a_\mu$. This model also has interesting signatures at muon colliders which will be discussed in the 
following.

\section{$\mu^+\mu^-\to \mu^+\mu^-, \tau^+\tau^-$ in 
$U(1)_{L_\mu - L_\tau}$ Models}

In this section we study experimental signatures of the 
$U(1)_{L_\mu - L_\tau}$ models at muon collider using the processes 
$\mu^+\mu^-\to \mu^+\mu^-, \tau^+\tau^-$. The Feynman diagrams which 
contribute to these processes are shown in Figs. 5 and 6. For the cases
a) and b) there are two new diagrams for $\mu^+\mu^-\to \mu^+\mu^-$
besides the SM ones, but
there is only one new diagram for $\mu^+\mu^- \to \tau^+\tau^-$ shown in 
Fig. 5. 
%
%
For case c),
there is no contribution from $Z'$ exchange for $\mu^+\mu^-\to \mu^+\mu^-$,
but there is one for $\mu^+\mu^- \to \tau^+\tau^-$ shown in Fig. 6. 
%

In Fig.~7 we show the cross section for $\mu^+ \mu^- \to \tau^+ \tau^-$
in case b), where each line represents constant $\Delta a_\mu$. 
We fixed $\sqrt{s} = 500$ GeV, and a corresponding 
SM cross section is
451.49 (fb).
There are $s$-channel photon, $Z$, $Z'$ and Higgs contributions. It
turns out that the Higgs contributions are negligible, if its mass is
far from $\sqrt{s}$. In our calculations we have used Higgs mass close
to the experimental lower bound.
When $\sqrt{s}$ is close to $m_{Z'}$, there is a resonance, even though
we have used a finite width $\Gamma_{Z'}$ calculated in the model
with 
$\Gamma_{Z'}=
\Gamma(Z'\to \nu_{\mu(\tau)}\bar{\nu}_{\mu(\tau)})+
\Gamma(Z'\to \mu(\tau)\bar\mu(\bar\tau))
$ 
because it is small.
We can clearly see the resonance effects. The cross section can be
enhanced quite dramatically compared to the SM. Therefore the muon colliders
can clearly show the new $Z'$ effects if case b) is realized in nature.

In Fig.~8 the cross section for $\mu^+ \mu^- \to \mu^+ \mu^-$ are shown in
case b). Since we neglected the muon mass, $t$-channel contribution shows
collinear singularity. We imposed angular cuts $ |\cos(\theta_{13})| <0.5$
when obtaining the total cross section. The corresponding cross section
for the SM is $1153.7$ (fb) for $\sqrt{s} = 500$ GeV.
In contrast to $\mu^+ \mu^- \to \tau^+ \tau^-$ process the cross
section does not decrease fast as $m_{Z'}$ increases due to the large
$t$-channel contributions.

In Fig.~9 we show the cross section of $\mu^+ \mu^- \to \tau^+ \tau^-$
as a function of $m_{Z'}$ for case c) for the
parameters which satisfy the $\Delta a_\mu$ constraint. 
Assuming the design luminosity 50 (fb$^{-1}$) per year,
we expect about 1000 deficit in the number of $\tau^+ \tau^-$
production events compared to the SM prediction.
We also note that this is also in contrast to the case b) where 
$m_{Z'}$ can be large and
the cross section can be highly enhanced compared to the SM case. 
Therefore we can see that the muon colliders can easily discriminate the
three different realization of $Z'$ models as well as the SM.

\section{Discussions and Conclusions}

In this paper we have shown that the gauged $U(1)_{L_\mu - L_\tau}$
models may contribute significantly to the anomalous muon magnetic dipole
moment. It is possible to produce a large enough value to account for the
discrepancy between in SM prediction in Ref.~\cite{dh,marciano}   
and experimental measurement from
BNL.
The relevant parameters are tightly constrained. The $Z'$ gauge boson
mass can vary from zero to large mass 
depending on
how the $U(1)_{L_\mu-L_\tau}$ manifest itself in nature. 
We have discussed three different cases. We find that in all
cases there are allowed parameters within which a large enough muon 
anomalous magnetic dipole moment given in eq. (1) can be generated.
In case a) the $Z'$ coupling parameter $a$ is restricted to be in the range 
$ 0.8\times 10^{-3} < |a| < 2.24\times 10^{-3}$.
In case b), the constraints on $a$ and $m_{Z'}$ are correlated. In the
heavy $Z'$ limit, $a/m_{Z'}$ is restricted to be in the range 
$9.2\times 10^{-3} <a/m_{Z'} ({\rm GeV}) < 25.8 \times 10^{-3}$.
In case c), the constraints on the $a$ and $m_{Z'}$ are even more restrictive.
The allowed $Z'$ mass is restricted to be around 1.2 GeV. 

In all the models discussed, 
the electron anomalous magnetic dipole moment is not affected by $Z'$
exchange because no $Z-Z'$ mixing was introduced. Were there $Z-Z'$
mixing, $\Delta a_e$ will also be affected. 
The $\tau$ magnetic dipole moment is constrained. We find that: In
case a), $\Delta a_\tau = \Delta a_\mu$;
In case b), in the limit of large $Z'$ mass $\Delta a_\tau \approx 
(m_\tau/m_\mu)^2\Delta a_\mu$;
And in case c), in the limit of large $Z'$ 
mass $\Delta a_\tau \approx \Delta a_\mu$.

Within the allowed parameter space, there are also other 
interesting predictions.
We have studied several signatures of these models at muon colliders.
At muon colliders there may be large effects
for processes $\mu^+\mu^- \to \mu^+\mu^-, \tau^+\tau^-$.
It is possible to distinguish the SM from different $U(1)_{L_\mu-L_\tau}$
models.
Future muon colliders can provide interesting clues
about these models. 

\acknowledgements   
PK is grateful to E. Stewart for useful discussions.  This work was
supported in part by National Science Council under the grants NSC
89-2112-M-002-016 and NSC 89-2112-M-002-062, in part by the Ministry
of Education Academic Excellent Project 89-N-FA01-1-4-3(XGH), in part
by DOE grant DE-FG03-96ER40969(NGD), and in part by by BK21 program of
Ministry of Education and SRC program of KOSEF through CHEP at
Kyungpook National University(PK).

\newpage
\begin{figure}
\begin{center}
\begin{picture}(280,220)
  \put(0,0){\epsfig{file=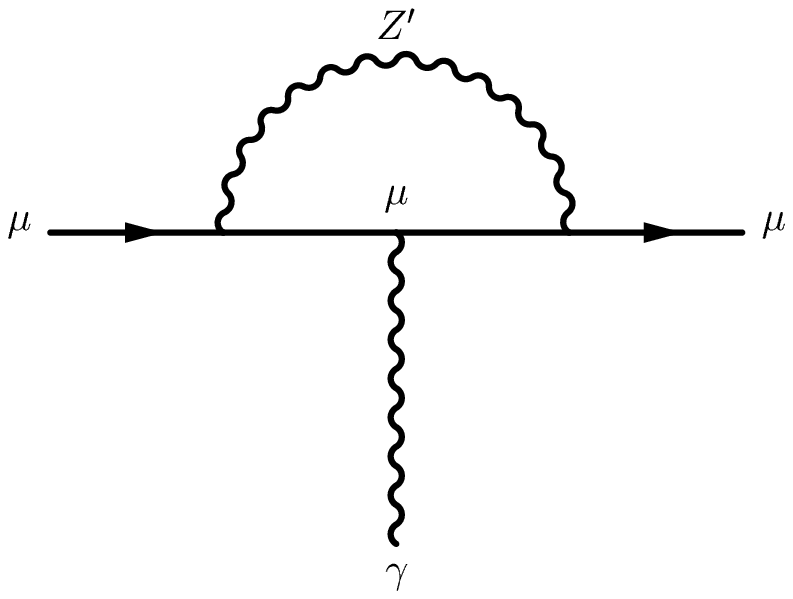,width=10cm,height=9cm}}
\end{picture}
\end{center}
\label{feyn_g2}
\caption{Feynman diagram which generates a non-zero $\Delta a_\mu$}
\end{figure}
\begin{figure}
\begin{center}
\begin{picture}(500,300)
  \put(0,0){\epsfig{file=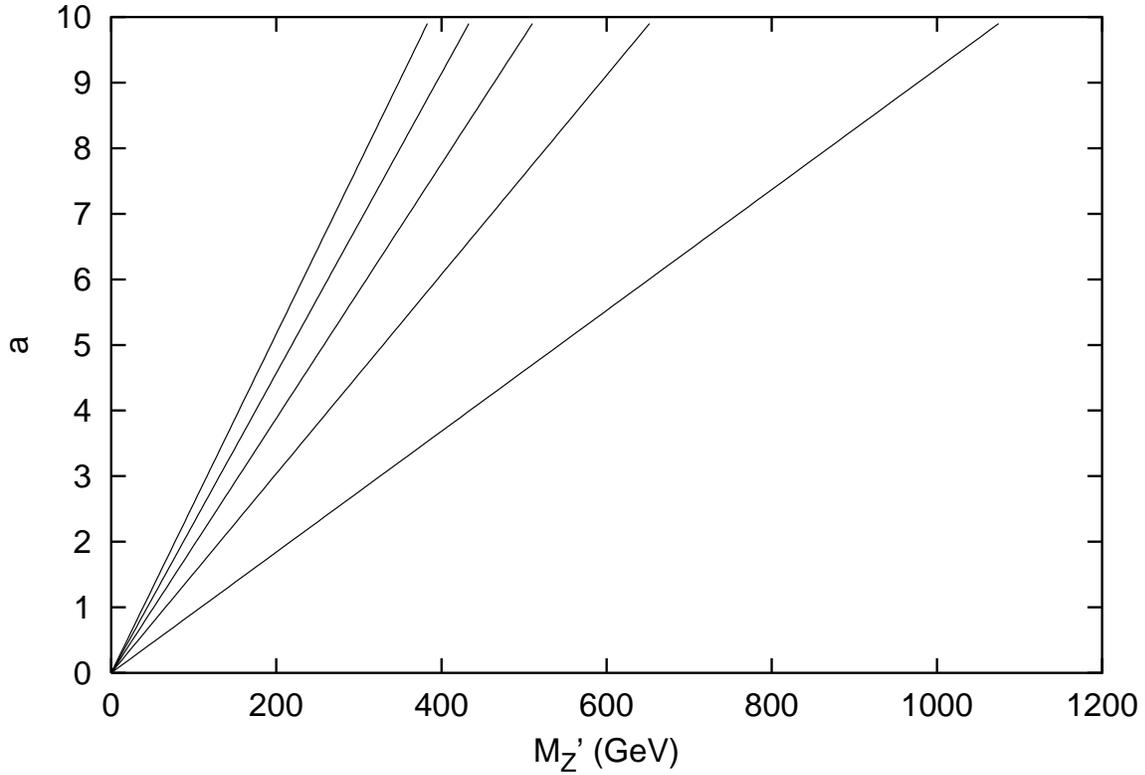,height=10cm}}
\end{picture}
\end{center}
\caption{$\Delta a_\mu$ on the $a$ vs. $m_{Z'}$ plane
in case b). The lines from left to right are for $\Delta a_\mu$
away from its central value at $+2\sigma,+1 \sigma,0,-1 \sigma$
and $-2\sigma$, respectively.
}
\label{fig:caseb1}
\end{figure}
\begin{figure}
\begin{center}
\begin{picture}(500,280)
  \put(0,0){\epsfig{file=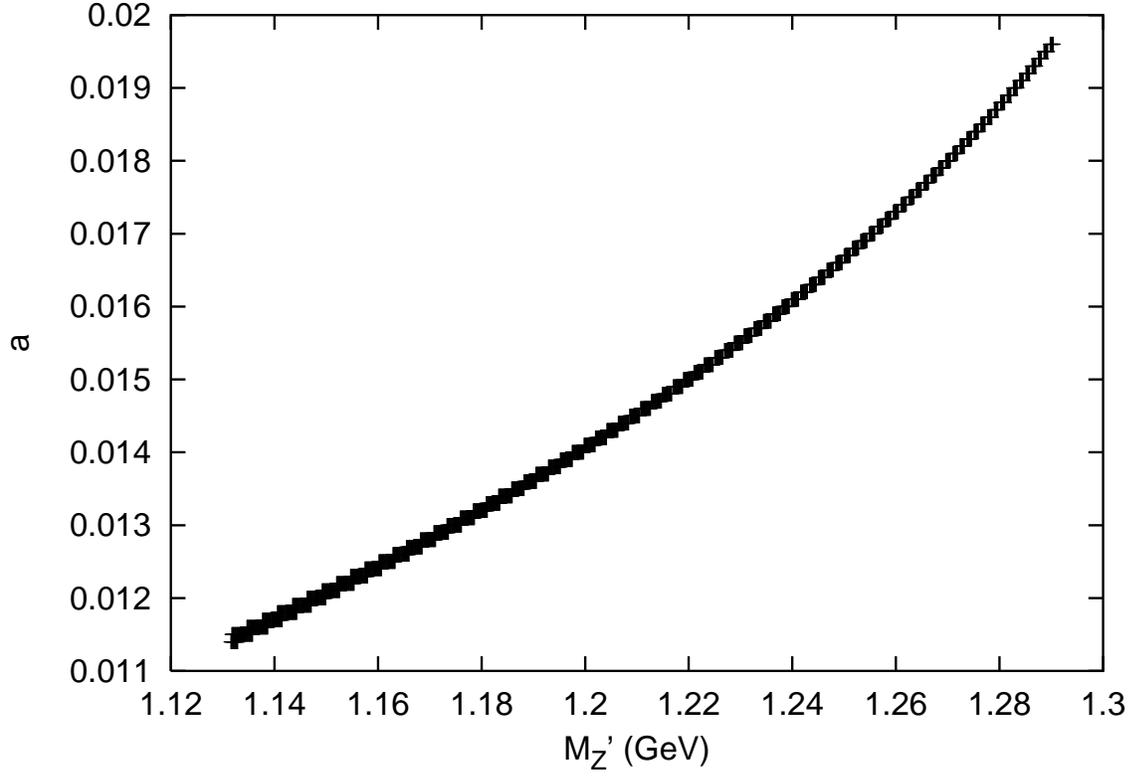,height=10cm}}
\end{picture}
\end{center}
\caption{The allowed region on the $a$ vs. $m_{Z'}$ plane 
with $\Delta a_\mu$ and $R$ varying in their
$2 \sigma$ allowed ranges for case c).}
\end{figure}
\begin{figure}
\begin{center}
\begin{picture}(500,280)
  \put(0,0){\epsfig{file=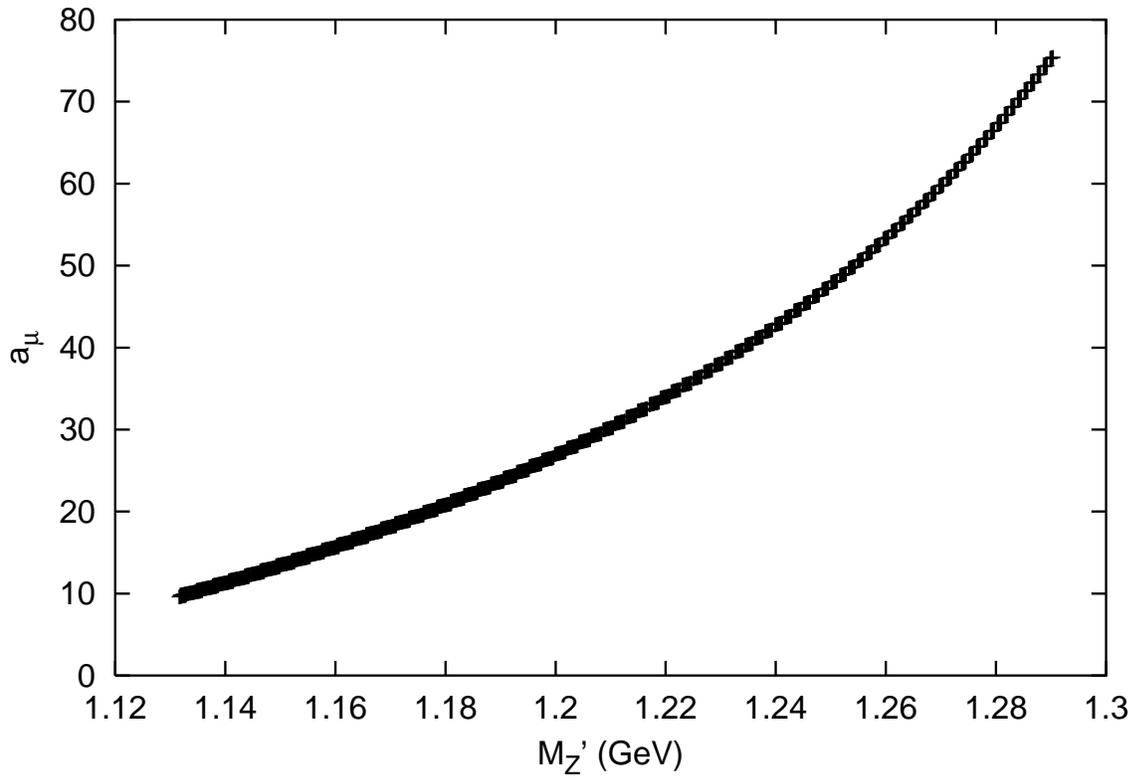,height=10cm}}
\end{picture}
\end{center}
\caption{$\Delta a_\mu$ in terms of $m_{Z'}$ for the
allowed parameters in the $R$ constraints in case c).}
\end{figure}
\newpage
\begin{figure}
\begin{center}
\begin{picture}(400,250)
  \put(0,0){\epsfig{file=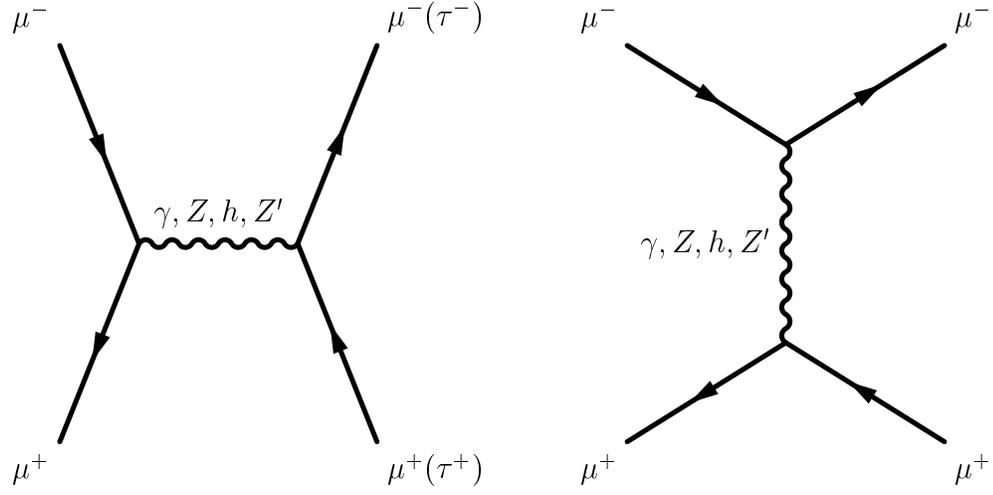}}
\end{picture}
\end{center}
\caption{Feynman diagrams for processes 
$\mu^+ \mu^- \to \mu^+ \mu^- (\tau^+ \tau^-)$ in case b)}
\end{figure}
\begin{figure}
\begin{center}
\begin{picture}(400,250)
  \put(0,0){\epsfig{file=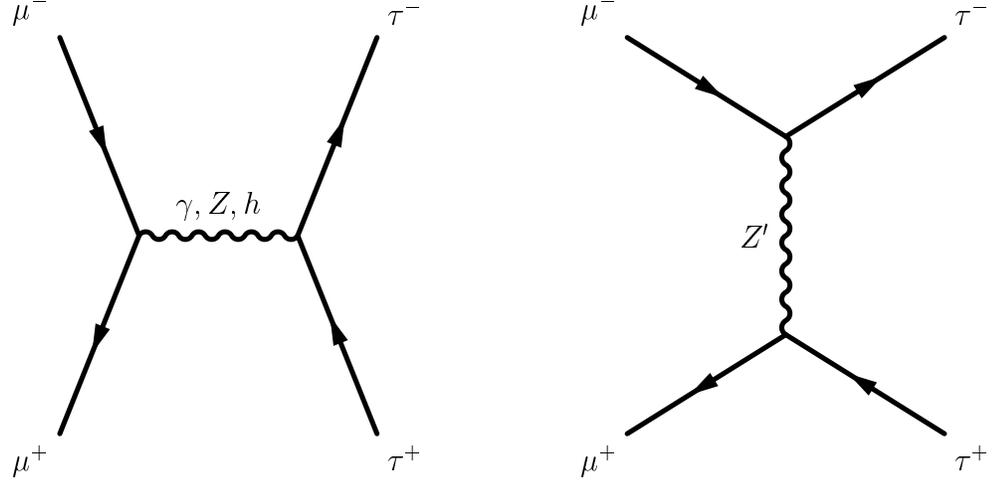}}
\end{picture}
\end{center}
\caption{Feynman diagrams for process
$\mu^+ \mu^- \to \tau^+ \tau^-$ in case c)}
\end{figure}
\newpage
\begin{figure}
\begin{center}
\begin{picture}(500,260)
  \put(0,0){\epsfig{file=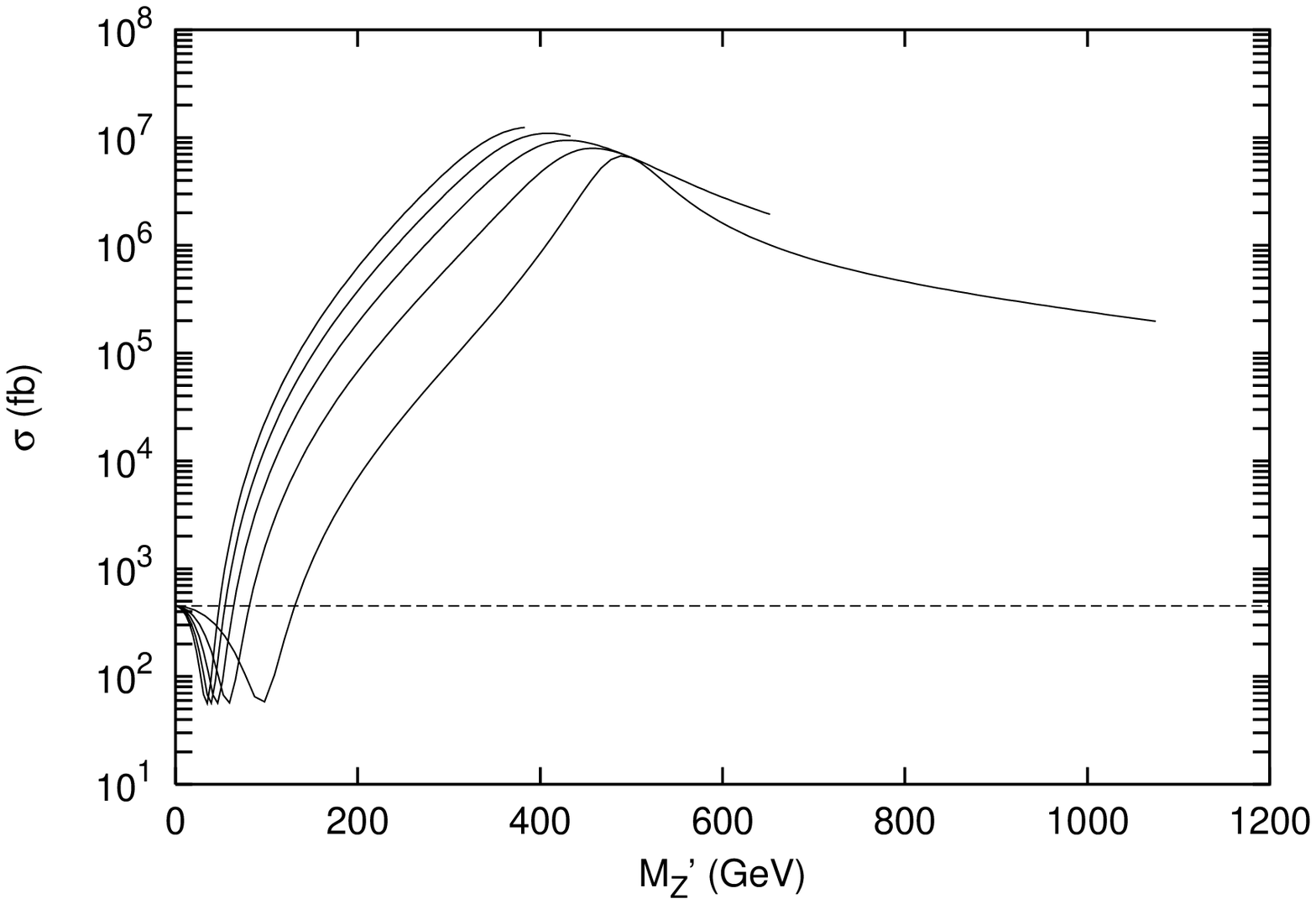,height=10cm}}
\end{picture}
\end{center}
\caption{Cross section for $\mu^+ \mu^- \to \tau^+ \tau^-$ 
as a function of $m_{Z'}$ for $\sqrt{s}=500$ GeV in case b). 
The lines from left to right correspond
to $\Delta a_\mu$ away from its central value at
$+2\sigma,+1 \sigma,0,-1 \sigma$ and $-2\sigma$, respectively.
The horizontal dashed line is the SM prediction.
}
\end{figure}
\begin{figure}
\begin{center}
\begin{picture}(500,260)
  \put(0,0){\epsfig{file=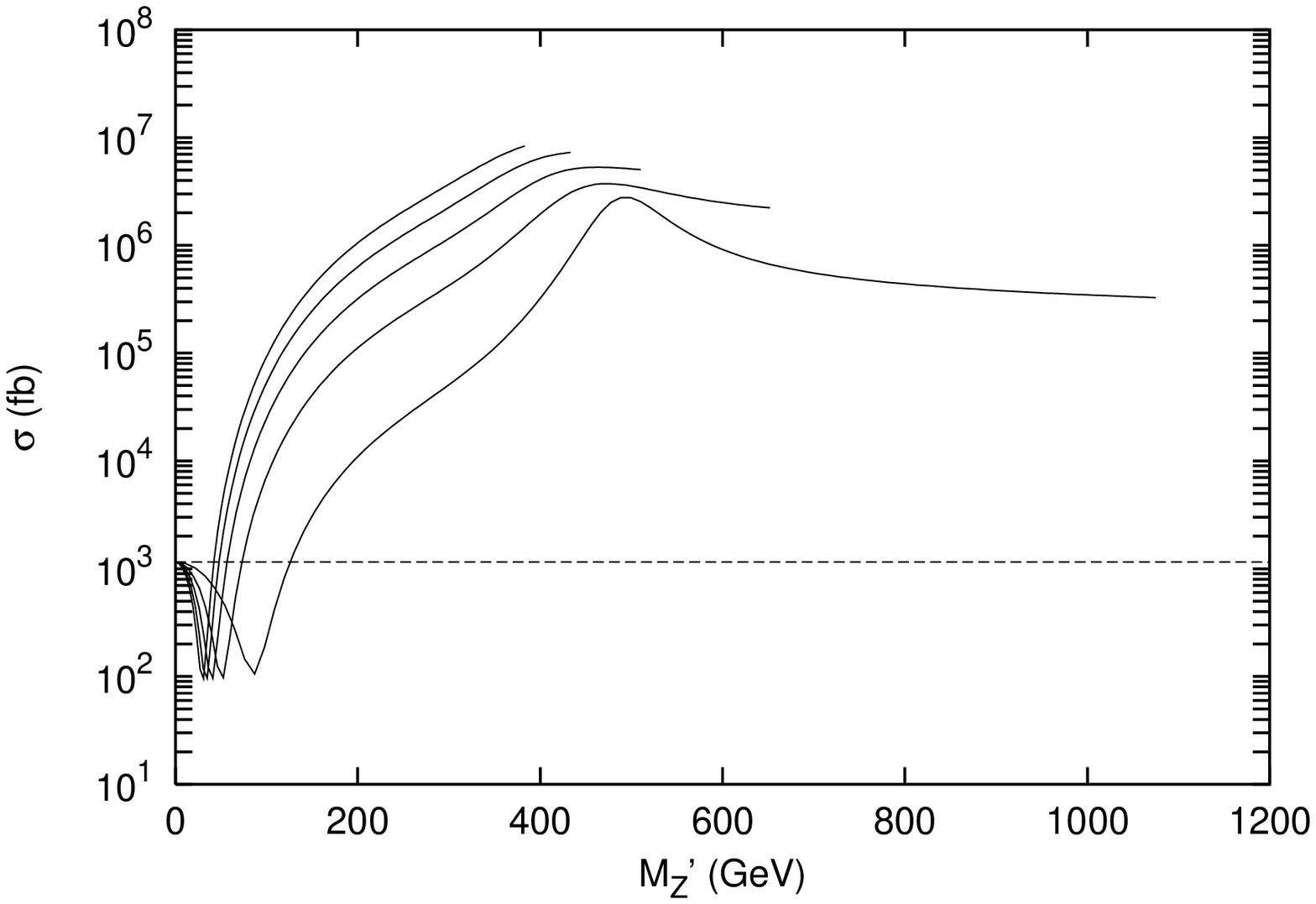,height=10cm}}
\end{picture}
\end{center}
\caption{Cross section for $\mu^+ \mu^- \to \mu^+ \mu^-$ 
as a function of $m_{Z'}$ for $\sqrt{s}=500$ GeV in case b). 
The lines from left to right correspond
to $\Delta a_\mu$ away from its central value at
$+2\sigma,+1 \sigma,0,-1 \sigma$ and $-2\sigma$, respectively.
The horizontal dashed line is the SM prediction.
}
\end{figure}
\newpage
\begin{figure}
\begin{center}
\begin{picture}(500,320)
  \put(0,0){\epsfig{file=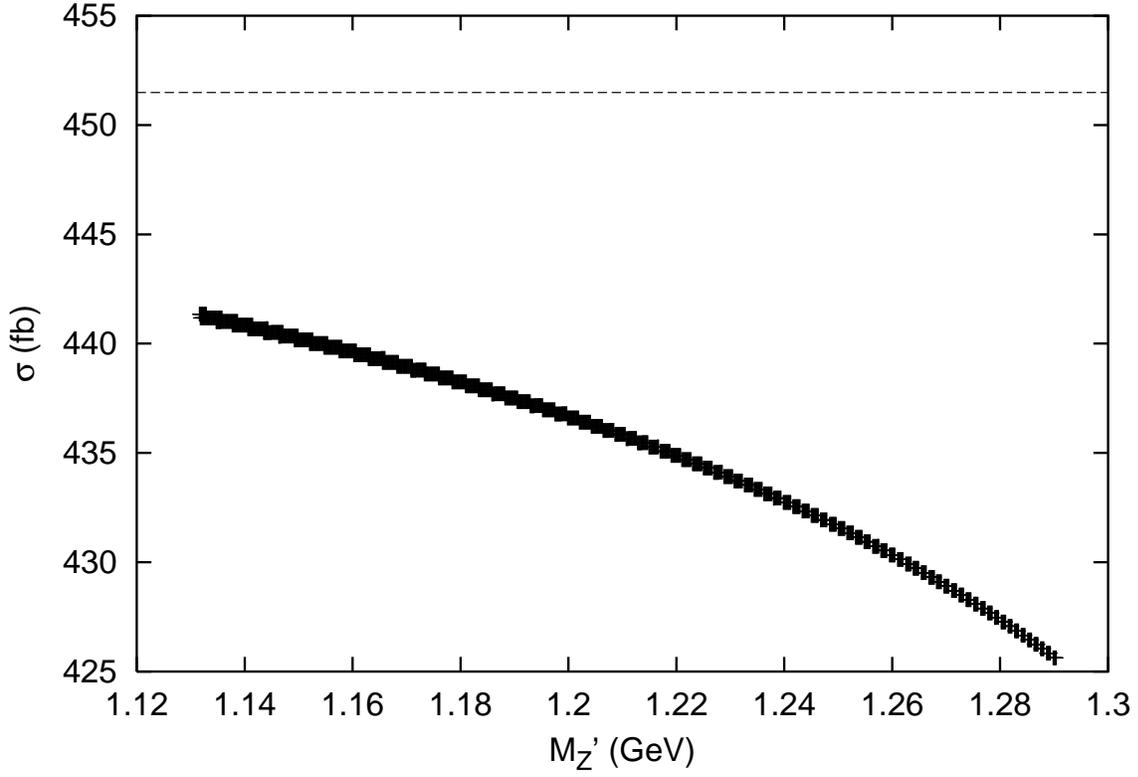,height=10cm}}
\end{picture}
\end{center}
\caption{Cross section for $\mu^+ \mu^- \to \tau^+ \tau^-$ 
as a function of $m_{Z'}$ for $\sqrt{s}=500$ GeV in case c).
The horizontal dashed line is the SM prediction. }
\end{figure}

\end{document}